# Statistical study of particle acceleration in the core of foreshock transients

Terry Z. Liu[1], Vassilis Angelopoulos[1], Heli Hietala[1], and Lynn B. Wilson III[2]

[1]Department of Earth, Planetary, and Space Sciences, University of California, Los Angeles, California, USA, [2]NASA Goddard Space Flight Center, Greenbelt, Maryland, USA

**Abstract**

Several types of foreshock transients upstream of Earth's bow shock possessing a tenuous, hot core have been observed and simulated. Because of the low dynamic pressure in their cores, these phenomena can significantly disturb the bow shock and the magnetosphere-ionosphere system. Recent observations have also demonstrated that foreshock transients can accelerate particles which, when transported earthward, can affect space weather. Understanding the potential of foreshock transients to accelerate particles can help us understand shock acceleration at Earth and at other planetary and astrophysical systems. To further investigate foreshock transients' potential for acceleration we conduct a statistical study of ion and electron energization in the core of foreshock transients. We find that electron energies typically increase there, evidently due to an internal acceleration process, whereas, as expected, ion energies most often decrease to support transient formation and expansion. Nevertheless, ion energy enhancements can be seen in some events suggesting an internal ion acceleration process as well. Formation conditions of foreshock transients are related to weak solar wind magnetic field strength and fast solar wind speed. Ion and electron energization are also positively correlated with solar wind speed.

1.      Introduction

The Earth's foreshock, which is upstream of the bow shock, is filled with back-streaming particles reflected at the bow shock or escaping from the downstream region [e.g., Eastwood et al., 2005; Burgess et al., 2012]. The foreshock is very dynamic; ULF waves and many types of foreshock transients have been observed and simulated. The latter include: hot flow anomalies (HFAs) [e.g., Schwartz et al., 1985], spontaneous hot flow anomalies (SHFAs) [Omidi et al., 2013; Zhang et al., 2013], foreshock bubbles (FBs) [Omidi et al., 2010; Turner et al., 2013; Liu et al., 2015], foreshock cavities [e.g., Sibeck et al., 2002], foreshock cavitons [e.g., Blanco-Cano et al., 2011], foreshock compressional boundaries [e.g., Omidi et al., 2009], and short large-amplitude magnetic structures (SLAMSs) [e.g., Schwartz et al., 1992]. Hot flow anomalies, spontaneous hot flow anomalies, foreshock bubbles and foreshock cavities have a common characteristic: a core populated by plasma with low field strength and low density, and associated with the presence of energetic particles and a deflection of the upstream flow away from the nominal solar wind direction (there is little deflection in foreshock cavities). The thermal pressure in the core is higher than that of the surrounding medium, which causes an expansion that can, if fast enough, create compressional boundaries and even shock waves on the core's outer edge. Because of the low density inside and the upstream flow deflection around the core, the core's dynamic pressure is very low. Thus, when foreshock transients with such a core encounter the bow shock, they cause it to move outwards, resulting in significant disturbances in the magnetosheath downstream and in the magnetosphere-ionosphere system [e.g., Sibeck et al., 1999; Turner et al., 2011]. Hot flow anomalies and foreshock bubbles require a solar wind discontinuity to concentrate foreshock ions and form a core. Their cores are larger (1-10 $R_E$) and hotter and are associated with larger plasma deflection than those of other foreshock transients. Thus, hot flow anomalies and foreshock bubbles can likely cause more significant effects on Earth's space environment [e.g., Archer et al., 2014; Archer et al., 2015].

In addition to the geoeffectiveness stemming from their dynamics, foreshock transients can accelerate particles [Kis et al., 2013; Wilson et al., 2013; Liu et al., 2016a; Wilson et al., 2016]. Because of its fast expansion against the solar wind, an FB can form a shock (an "FB shock") upstream of its core. Like the bow shock, an FB shock can accelerate solar wind particles through shock drift acceleration [Liu et al., 2016a]. Wilson et al. [2016] reported that electrons inside the cores of foreshock transients can be accelerated to hundreds of keV. Energetic particles accelerated by foreshock transients can be transported towards the Earth, affecting the magnetosphere-ionosphere system. Moreover, elucidating the acceleration potential of foreshock transients may also help us understand shock acceleration mechanisms. Shock acceleration at astrophysical shocks, a main source of energetic particles in the universe, have not been fully understood, including the so-called "injection problem" [Jokipii, 1987] (the required initial energization of particles prior to their eventual shock acceleration [e.g., Lee et al., 2012; Raymond et al., 2012]). Foreshock transients have not yet been fully integrated into models of particle acceleration at shocks. They hold high potential for providing the source for parent shock acceleration and affect parent shock structure, requiring further investigation which could improve our understanding of shocks.

Specifically, how are electrons accelerated, what controls the acceleration intensity, and what role do ions play in the core? In this paper, we statistically examine the characteristics of particle acceleration by foreshock transients. We address the role of ions, how frequently acceleration can occur, and what parameters control acceleration intensity. In addition, we discuss formation conditions of these foreshock transients.

In this study, we focus only on the acceleration in the core. We ignore energetic particles outside foreshock transients, such as those reflected and accelerated by an FB shock [Liu et al., 2016a], because in a statistical study it is difficult to distinguish whether such particles are reflected by foreshock transients or the parent bow shock itself.

In section 2, we introduce our database and explain how we assembled our event list. We also explain how we quantified the energization of ions and electrons. In section 3, we present our statistical results of ion energization and discuss the role of ions. In section 4, we show our statistical results of electron energization. In section 5, we address conditions for formation of foreshock transients. In section 6, we summarize and discuss our findings.

**2. Data and Methods**

Our database consists of measurements by THEMIS mission probes TH-B and TH-C in 2008 and 2009 [Angelopoulos, 2008]. In the first two dayside seasons of that mission [Sibeck and Angelopoulos, 2008], TH-B had an ~30 $R_E$ apogee upstream on the dayside, and TH-C had a lower apogee, ~20 $R_E$. We analyzed plasma data from the THEMIS electrostatic analyzer (ESA) [McFadden et al., 2008] and its solid state telescope (SST) [Angelopoulos, 2008] and magnetic field data from its fluxgate magnetometer [Auster et al., 2008]. We restricted our study to data in Fast Survey mode and Particle Burst mode to benefit from the higher angular resolution of particle data (listed in the supplementary material) and higher time resolution of magnetic field data (0.25s) in those modes. To obtain pristine solar wind data, we used OMNI 10 min averaged data.

By examining TH-B and TH-C observations upstream of the bow shock, we established an event list of 247 foreshock transients with a core associated with foreshock ions. They are listed in the supplementary material. Our selection criteria were as follows: the density inside the core should drop by more than 30% from that of the background or the pristine solar wind; the event should last longer than 10 s; plasma deflection (of more than 5%) should be observed. Thus, our events can include hot flow anomalies, spontaneous hot flow anomalies, foreshock bubbles, and those foreshock cavities that have a finite flow deflection. We exclude SLAMS and foreshock cavitons, however, as they are more related to waves and have different characteristics. Figure 1 is an example (an HFA) of an event in our list observed by TH-C.

In this study, we only consider particle energization inside the core of foreshock transients. The boundary of the core is defined as the point where density and field strength start to drop (two dashed vertical lines in Figure 1). The average value of the quantities described below is calculated in the time interval of the core. We also choose a background time interval within 5 min of the event either downstream or upstream depending on which region contains background foreshock particles (two dotted vertical lines in Figure 1) to get the average value for comparison.

To measure the energization intensity, we need to process the particle data beyond the standard moments and spectra. For ions, we first combine ESA and SST distributions. Next, we should calculate the first order moment to obtain mean kinetic energy and the second order moment to obtain mean thermal energy of the non-solar wind component to characterize ion energization. Thus, we need to remove the solar wind beam from the ion distribution before computing the moments. We obtain the rough solar wind velocity from the total ion distribution by confining the phi angle, and then remove counts in the bins that are less than 3-5 angular and energy bins away from the solar wind velocity in the ion distribution. Then we fill these bins by interpolation. Because the ESA treats all ions as protons (mass over charge = 1), an alpha beam from the solar wind (mass over charge = 2) will appear at energy around two times the solar wind proton energy. We removed the alpha beam in the same way as the solar wind protons. Figure 1e shows the energy spectra of the non-solar wind component after ESA and SST have been combined. It is evident that the typical solar wind beam around 1 keV (see Figure 1d) has been removed. After this, we calculate the density, mean thermal energy, mean kinetic energy, shown in Figure 1f, g, h, and other products. Inside the core, the density of the non-solar wind component increased, because to form an HFA, foreshock ions need to be concentrated by a driver discontinuity [e.g., Liu et al., 2016b], and some cold solar wind ions can be thermalized in the core and mix with the non-solar wind ions [Wang et al., 2013; Chu et al., 2017]. In Figure 1g, h, we see that inside the core, the non-solar wind ion mean thermal energy increased while the mean kinetic energy decreased. This is because after foreshock ions pass through a discontinuity, their parallel speed needs to be projected on a new direction and thus decreases (under certain IMF configurations, the perpendicular speed can also make a negative contribution to the parallel speed [Liu et al., 2015]). In the discontinuity frame where the convection electric field is zero, particle energy is conserved. The decreases in kinetic energy (calculated from mean parallel speed) result in increases in thermal energy [e.g., Archer et al., 2015; Liu et al., 2015]. The increased thermal pressure of the core causes the expansion of foreshock transients.

In addition to these moments, we calculate the Kappa index of ion phase space density distribution as another energization estimate by assuming that ion distributions follow a Kappa distribution. A smaller Kappa index means a stronger deviation from thermal equilibrium. To fit the Kappa distribution, we transform the distribution of non-solar wind ions into their rest frame and obtain the corresponding energy spectra (methods of frame transformation and energy spectra calculation are described in the supplementary material).

As for electrons, we combine ESA and SST electron distributions only in Particle Burst mode; we do not combine them in Fast Survey mode, because of the poor angular resolution. Because this may cause statistical differences, events in different modes are labeled separately in Section 4. Because we cannot directly remove solar wind electrons from the distribution, we only concern ourselves with their suprathermal tail. We define the suprathermal component as that with energy greater than three times the average temperature (Figure 1j, blue line). To quantify the electron energization, we calculate the average temperature of the suprathermal tail (Figure 1j red line, from ESA only in Fast Survey mode and from combined distribution in Particle Burst mode). We also calculate the partial pressure of the suprathermal tail, and then we calculate $E_{5\%}$, the energy above which the bins contribute 5% to the partial pressure (in Fast Survey mode, we calculate partial pressure in ESA and SST separately, and then add them up; in Particle Burst mode, we calculate it from combined distribution). As with ions, we also calculate the Kappa index of the electron phase space density distribution below 1 keV (to avoid noise) after the frame transformation into the electron rest frame (Figure 1k). To approximate the maximum energy of electrons, we also determine the maximum energy bin for the ESA and SST, defined as the highest energy bin of either instrument that can observe fluxes above the 1-count level after instrumental background noise subtraction. When determining the maximum energy in the SST energy range in Fast Survey mode, we also require that the electron signal be detected by at least the lowest three energy channels to fully distinguish it from noise.

## 3. Statistical result of ion energization

After we obtain the non-solar wind ions' mean kinetic energy, mean thermal energy, and Kappa index (described in Section 2), we can correlate their statistical characteristics and compare them to the solar wind.

We first compare the mean total energy (mean kinetic energy plus mean thermal energy) inside the core with the value in the background foreshock. As shown in Figure 2, the mean total energy inside the core overall is a little smaller than that in the background foreshock. So why is the ion energy decreased? One possible reason is expansion of the foreshock transients. Because of the high thermal pressure in the core, the core expands and pushes the surrounding solar wind cold plasma and magnetic field. This process requires ions to do work approximately equal to the volume change multiplied by the pressure. To further explore this scenario, we identified all events with a shock-like structure (i.e., pink circles in Figure 2). Almost all such events are below the diagonal in Figure 2. We suggest that this is because the amount of energy ions lose depends on the speed of expansion. If a shock is observed, the expansion must be faster than the local fast-mode speed. Therefore, more of the plasma's internal energy needs to be consumed for the expansion. The ion energy may also be used to generate waves. However, wave energy can be much smaller than the expansion energy which is comparable to the solar wind kinetic energy. Another possible reason is that some remaining cold solar wind ions can be thermalized in the core [Wang et al., 2013; Chu et al., 2017] and mix with the non-solar wind ions. As the energy of a solar wind ion is generally smaller than the energy of a foreshock ion, the mixing of thermalized solar wind ions could result in an underestimate of the energy of non-solar wind ion component.

Although ions must lose energy to support the formation and expansion of foreshock transients and their energy are probably underestimated, ~30% of our events have a mean total energy inside the core that is larger than the value in the background foreshock. Thus, there could be ion acceleration processes inside the core. One possible zero-order acceleration mechanism inside the core is reflection at the earthward-moving upstream boundary of foreshock transients. Let us consider foreshock bubbles as an example. Foreshock bubbles and their upstream boundary should convect with the solar wind toward the bow shock. Reflection in the core at its upstream boundary will give ions additional velocity approximately equal to two times the normal speed of the upstream boundary in the Earth frame. Foreshock bubbles also have an earthward-moving downstream boundary, however. Because FBs expand, the downstream boundary should move faster towards the bow shock than the upstream boundary (in the Earth frame). Therefore, ions trapped in the core between these two boundaries will lose more energy by reflecting at the downstream boundary than they would gain by reflecting at the upstream boundary. This is a picture of energy transfer from ions to expansion from the perspective of a single ion. However, after FBs connect to the bow shock, the downstream boundary also merges with the rest of the bow shock. Remaining ions and those that are later trapped in the core could gain energy during this time interval. Higher ion energy than that in the background foreshock can thus be observed if the energy gain during this process is larger than the energy loss.

Next, we examine how the solar wind speed affects non-solar wind ion energy. Figure 3a, b, shows that non-solar wind ion mean kinetic energy and mean thermal energy inside the core increase with solar wind kinetic energy. We can also see that the Kappa index decreases with increasing solar wind kinetic energy (Figure 3c). Thus, the solar wind speed can dramatically affect the ion energy. This is probably because faster solar wind speed can give higher foreshock ion energy, which is the initial energy of particle source for the foreshock transients. Additionally, mixed thermalized solar wind ions may make a partial contribution to the correlation between the mean thermal energy and the solar wind kinetic energy (Figure 3b).

If we plot the ratios of our energization estimates inside the core over those in the background foreshock (the normalized energization estimates) against the solar wind kinetic energy (Figure 3d, e, f), we can see that the correlation that held for the un-normalized estimates disappears except for the Kappa index. There are two possible explanations for this lack of correlation. The first is that high-energy ions could leak out from foreshock transients and affect the ion energy in the background foreshock. The gyroradii of ions with an energy of several keVs can be 1000s of km, i.e., comparable to the boundary thickness of foreshock transients (100s – 1000s of km). Once foreshock ions are accelerated to several keV or tens of keV, they could easily leak out and increase our estimated background ion energy. This effect may also help explain why ion energy inside the core is lower than that in the background foreshock in most of our events. The second possible explanation for the lack of correlation may be that the ion energy loss in the core is related to (proportional to) the transient's expansion speed, which, in turn, was shown previously [Liu et al., 2016b] to be proportional to the solar wind speed. At the same time, the initial ion energy is also proportional to the solar wind speed. Thus, this may prevent the ratio of these two quantities from exhibiting a clear correlation to the solar wind speed.

**4. Statistical results of electron energization**

Now we analyze the statistical results of electron energization (see Figure 4). Calculating the average temperature of the suprathermal tail and $E_{5\%}$, the energy above which the bins contribute 5% to the partial pressure, we found that nearly 90% of the events have core electrons that are more energized than the ones in the background foreshock (see Figure 4d, e, horizontal lines). More than 70% of events also show a smaller Kappa index inside the core than that in the background foreshock (see Figure 4f, horizontal line), indicating stronger energization. Thus, most foreshock transients energize electrons. In ~30% of our events, the maximum energy is larger than 25 keV (minimum SST energy range) in Figure 4g. In some of the events, energies are hundreds of keV to 700 keV, consistent with the 100-300 keV events reported by Wilson et al. [2016] (their events are shown in green in Figure 4g).

Electron energy estimates in Fast Survey mode (black) and Particle Burst mode (orange) show no clear statistical differences except at around 25 keV in Figure 4g. This is because in Fast Survey mode SST energy range, we require electrons to be detected by at least the lowest three energy channels, i.e., the threshold for the electron maximum energy is ~50 keV.

Next, we examine whether electron energization depends on solar wind speed. We find that average temperature of the suprathermal tail (Figure 4a) and $E_{5\%}$ (Figure 4b) increase with solar wind kinetic energy. Additionally, electron Kappa indices (Figure 4c) decrease with increasing solar wind kinetic energy. Comparing the maximum energy that can be detected by the particle instruments with the solar wind speed, we see that the highest maximum energies more likely occur at high solar wind speed (Figure 4g, SST energy range). Normalizing our energization estimates in the core to those in the background foreshock (Figure 4d, e, f), we find that the correlations (except for the Kappa index) still hold, especially when the solar wind kinetic energy is below ~2 keV (~620 km/s). This is different from our ion results. This means that the solar wind speed can not only provide higher initial energy, but can also affect the electron acceleration process. This effect, however, appears to saturate when the solar wind speed becomes too large (> 2 keV, ~620 km/s in Figure 4).

How, then, could foreshock transients accelerate electrons? One possible way is the reflection at the upstream boundary like ions [e.g., Omidi et al., 2010]. Because electrons are fast, they can bounce between boundaries multiple times. If this bouncing is between the bow shock and the earthward-moving upstream foreshock boundary, electron energy could increase through Fermi acceleration. This mechanism, however, is debated. For example, if the electrons are too energetic (e.g., 100s of keV), their very large gyroradii (~100s – 1000s of km) could easily cause them to leak out. Other possible mechanisms may be related to waves in the core, especially whistler waves. More case studies and simulations are needed to reveal details of electron acceleration mechanisms.

## 5. Formation conditions of foreshock transients

We have addressed the potential dependence of ion and electron energization at foreshock transients on solar wind speed, but not yet the potential effects of other solar wind parameters such as density, temperature, and magnetic field strength. We found these solar wind parameters to be correlated with energization estimates. Because the fast solar wind is hot and tenuous and the slow solar wind is cold and dense, however, we must attempt to isolate the effects of the natural characteristics of the solar wind. Towards that end, we compare the solar wind parameters of our foreshock transient events to those from the entire time interval of our database. If those typical parameters differ in the two datasets, the formation condition of foreshock transients could be revealed.

We obtain the ambient solar wind characteristics by using 10 min OMNI data from 2008-06-16 to 2008-11-08 and from 2009-06-27 to 2009-09-05, the entire time interval of our database. By comparing Figure 5a with 5b, and 5c with 5d, the relationship between the solar wind density and temperature against solar wind speed for our events is almost identical to that in the generic solar wind from entire database. However, the solar wind magnetic field strength for our events (Figure 5e) is almost along the lower bound of the solar wind magnetic field strength measured in the entire time interval of our database (Figure 5f). Thus, weak magnetic field strength favors formation of foreshock transients, probably because foreshock transients can expand easier and be more dynamic in the weaker solar wind field strength.

If we look at the distribution of solar wind speeds (Figure 6), we notice that in the ambient solar wind ensemble, 55% of the data are in the slow solar wind category (smaller than 400 km/s) and the relative portion of the distribution decreases with increasing solar wind speed (Figure 6a). However, the solar wind speed distribution of the foreshock transient events shows that only 26% are in the slow solar wind category (Figure 6b). If we calculate the ratio of two distributions (Figure 6c), we can see that the probability of foreshock transient formation increases with solar wind speed. We interpret this as evidence that a higher solar wind speed can provide more energy for formation of foreshock transients. Therefore, fast solar wind speed is another condition that favors formation of foreshock transients. This result is consistent with findings that HFAs are favored by the fast solar wind [Chu et al., 2017].

**6. Summary and discussions**

We have shown that the most important role of energetic (heated) ions in the core is to enable formation and expansion of foreshock transients, which create an environment for particle acceleration. Even though ions must give up some of their energy so that foreshock transients can fully form and expand, net energization over and above the background is ultimately occurring when the core collapses onto the bow shock. The net energization can be observed occasionally (30%) for ions, and almost always (90%) for electrons. Thirty percent of foreshock transients have maximum electron energies of more than 25 keV.

Solar wind speed is the parameter having the highest positive correlation with (and likely the dominant factor controlling) the energization. It is positively correlated with all our ion and electron energization estimates, likely because higher solar wind speed can provide a particle source of higher initial energy. The ratio of our electron energization estimates (except the Kappa index) in the core to the same estimates in the background foreshock is also correlated with solar wind speed, suggesting that solar wind speed can also increase electron acceleration efficiency. By comparing other solar wind parameters during our transient events with the same parameters computed from the entire time interval of our study, we found that weak magnetic field strength and fast solar wind speed favor formation of foreshock transients.

Next, we discuss the role of ions and energy transfer during the lifetime of a foreshock transient using HFAs and FBs as examples. Foreshock ions first encounter and pass through the solar wind discontinuity responsible for HFA or FB formation. During this process, part of their kinetic energy is converted into thermal energy. Hot, concentrated foreshock ions create a high thermal pressure core that starts to expand. During the expansion, the core's increasing size means that work is done to the surrounding cold, solar wind plasma. Therefore, the overall energy of concentrated ions in the core should decrease. Some of these ions are expected to leak out, some to remain inside, and new foreshock ions are expected to continue being trapped from outside and lose some energy to support the expansion. Because the expansion speed is normally smaller than the solar wind speed, the upstream boundary should move towards the bow shock at a speed slower than the solar wind speed. In the spacecraft frame (Earth frame), ions in the core will gain energy by reflecting at the moving upstream boundary. However, the reflection at the downstream boundary, which moves towards the bow shock faster than the upstream boundary, will make them lose more energy than they gain by reflection at the upstream boundary. This is another way to explain how overall ion energy decreases. When the foreshock transient starts to connect to the bow shock, the volume of the core starts to decrease. Both ions and electrons no longer reflect and lose energy at an earthward-moving downstream boundary; that boundary is now merged into/transmitted through the bow shock. Rather, they will reflect between a fixed or even sunward-moving boundary, the bow shock, and an approaching upstream boundary of the foreshock transient, and will gain energy. This process is to release energy that had been stored during the foreshock transient formation and expansion process from the early thermalized, concentrated ions. Therefore, as far as particle energization to suprathermal energies is concerned, the most important role of ions is to establish the structure of foreshock transients and store their energy inside the structure so it can later-on accelerate electrons and some ions as it collapses onto the bow shock.

Almost all foreshock transients convect with the solar wind and finally merge into the bow shock. Electron energy increases were observed in more than 90% of our events. However, since both electrons and ions can gain energy during this process (by reflecting at the earthward-moving upstream boundary and the bow shock), it is not immediately obvious why ions do not exhibit energization in most of our events.

One possible reason is that electrons move much faster than ions, so electrons can be accelerated through many more reflections than ions during the same timescale (e.g., 1 min) of the foreshock transient's collapse. Electrons could also be accelerated through some other mechanisms, e.g., by whistler mode waves. Thus, we hypothesize that the ion energy increase may not be strong enough to compensate for the energy loss during the expansion process. The second possible reason is the larger gyroradii of energetic ions which make them more likely to leak out of the core compared to electrons of a comparable energy. This leakage will lower the energy in the core and increase the energy in the background foreshock. (The leaked electrons, on the other hand, can move quickly away from the local neighborhood of the foreshock transient, and thus do not have the opportunity to change the average electron energy in that neighborhood.) The third possible reason is that we could be underestimating the ion energies in the core due to mixing with some thermalized solar wind ions.

We have shown that the solar wind speed is the most important parameter that affects energization. Now we discuss the role of the solar wind speed. With regards to ions, higher solar wind speed can result in higher ion energy, probably because the solar wind speed can increase the initial energy of the particle source (background foreshock ions). However, when normalized to the energization estimates of the background foreshock, the core energization does not correlate well with the solar wind speed. One possible reason is that high energy ions leak out into the background foreshock. Another reason may be that both the energy loss due to expansion and the initial energy are proportional to the solar wind speed, so their ratio can hardly exhibit a clear correlation with the solar wind speed. With regards to electrons, both the electron energization estimates and their normalized counterparts are correlated with the solar wind speed. Thus, the solar wind speed can increase not only the initial energy of particle source but also the efficiency of the electron acceleration process.

But how does solar wind speed affect the acceleration process? One candidate acceleration mechanism is Fermi acceleration [e.g., Omidi et al., 2010], in which electrons bounce between converging boundaries: the earthward-moving upstream boundary of the foreshock transient and the bow shock. In this case, the electron energy is related to the initial distance between the two boundaries, i.e., the size of the foreshock transient when it connects to the bow shock. Because a faster solar wind speed can make the transients expand further, resulting in higher compression between walls during the transient's collapse, this is one possible way for the solar wind speed to affect the electron acceleration process. To further investigate this scenario and reveal the exact acceleration mechanism, case studies and simulations need to be applied in future.

Although shock acceleration is one of the most important acceleration mechanisms in the universe, we do not fully understand it. For example, diffusive shock acceleration at shocks requires particles that have large initial energy, but where such particles obtain this energy is still unknown (the so-called "injection" problem [Jokipii, 1987]). We have shown that foreshock transients often pre-accelerate particles before they interact with the shock and bring them towards the bow shock supporting previous results [e.g., Liu et al., 2016a; Wilson et al., 2016]. Foreshock transients could in principle form everywhere throughout the universe upstream of shocks, and thus they may be quite important for addressing the injection problem. Further studies of the potential of foreshock transients for particle acceleration may help solve shock problems in general.

In this study, we also showed that the formation of foreshock transients is favored by the weak solar wind magnetic field strength and fast solar wind speed.

This result can help us in the future to establish a model to predict the formation of foreshock transients. Studies of the formation conditions of foreshock transients at other planets, such as Mars and Saturn, and simulations can further advance our knowledge on how foreshock transients are formed.

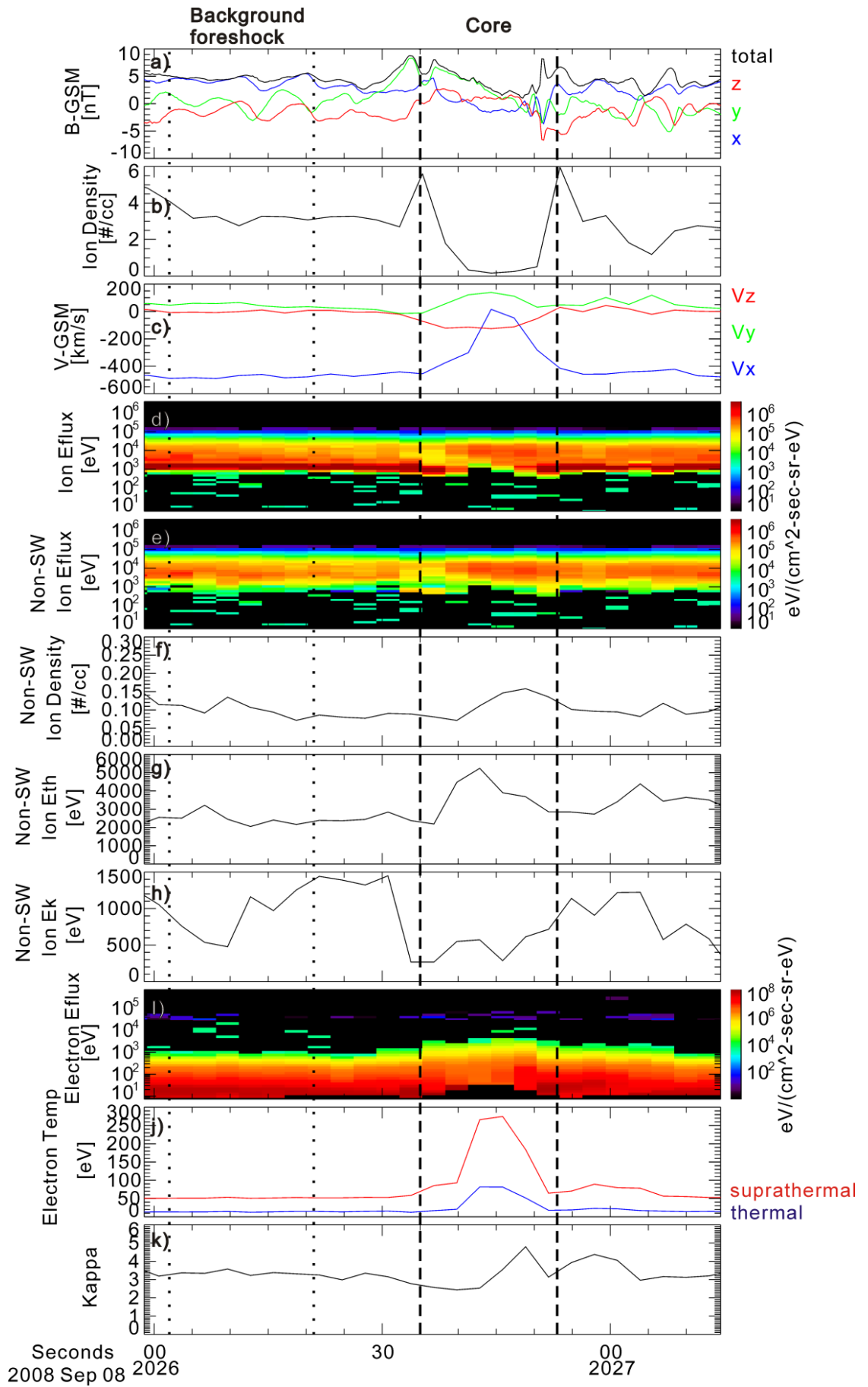

Figure 1. TH-C observation of a typical foreshock transient (an HFA) and various data products constructed from the particle distributions and used further in our statistical study. From top to bottom: (a) magnetic component in GSM coordinates (XYZ, total in blue, green, red, black, respectively); (b) total ion density; (c) ion bulk velocity in GSM coordinates (XYZ in blue, green, red, respectively); (d) ion ESA and SST (combined) energy spectra; (e) ion ESA and SST (combined) energy spectra of non-solar wind component; (f) ion density of non-solar wind component; (g) ion mean thermal energy of non-solar wind component; (h) ion mean kinetic energy of non-solar wind component; (i) electron ESA and SST (combined) energy spectra; (j) electron average temperature of total population (thermal, in blue) and suprathermal tail (in red); (k) Kappa index of the electron phase space density distribution in the energy range below 1 keV.

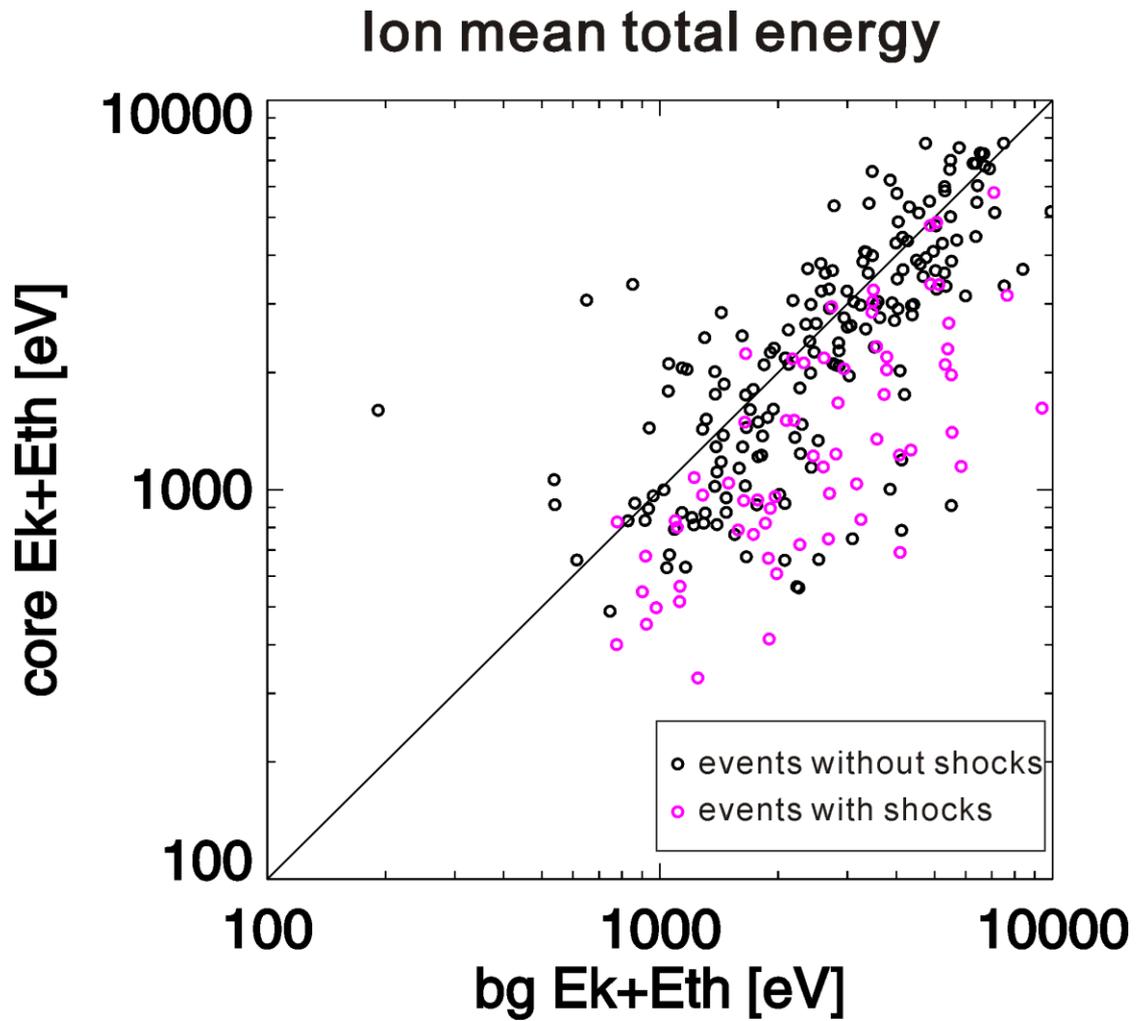

Figure 2. Non-solar wind ion mean total energy in the core vs in the background foreshock. Pink indicates events with a shock-like structure. 70% of events have ion mean total energy smaller than that in the background foreshock (circles below the diagonal). Almost all the events with a shock like structures (pink circles) are below the diagonal.

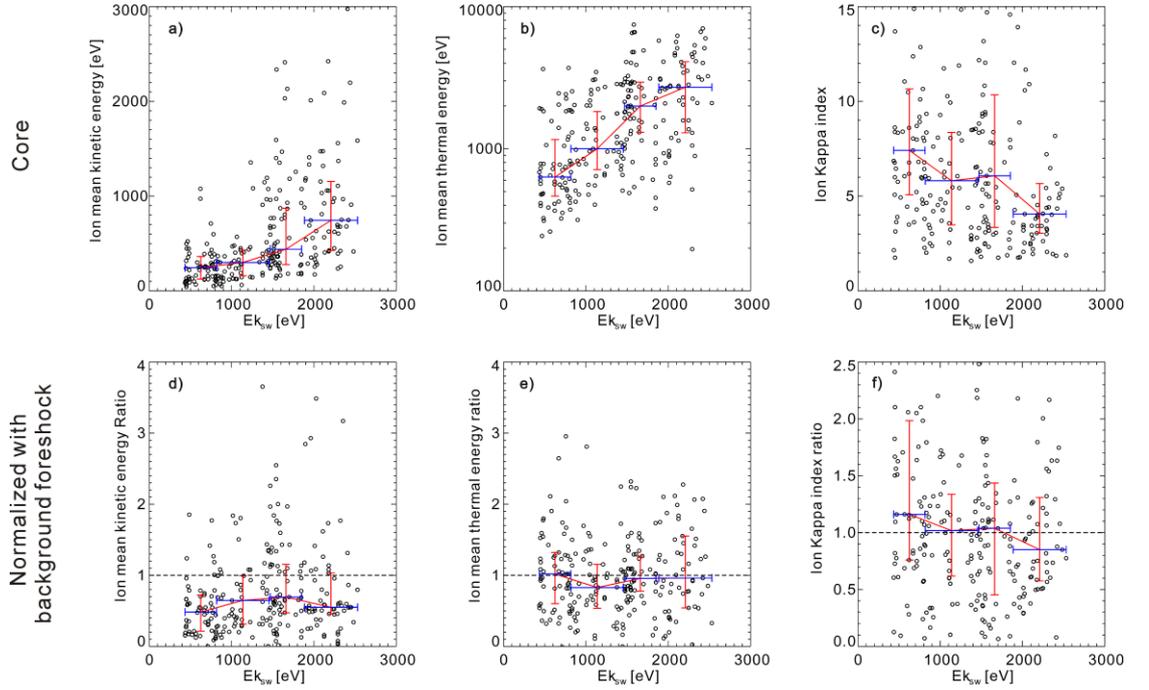

Figure 3. Non-solar wind ion energization estimates in the foreshock transient core plotted against the solar wind kinetic energy: mean kinetic energy (a); mean thermal energy (b); Kappa index of the phase space density distribution (c). Normalized estimates (d, e, f, respectively) are the ratios of these quantities to the same quantities in the background foreshock. Horizontal dashed line in (d, e, f) is at ratio = 1. Red line indicates the median and quartiles (25% and 75%). Dark blue horizontal bars indicate the data bins in which median and error bars are calculated (each data bin contains equal data number). Positive correlation between ion energization in the core and the solar wind speed can be seen in (a, b, c), but the correlation is barely seen after normalizing mean kinetic energy and mean thermal energy to those in the background foreshock (d, e).

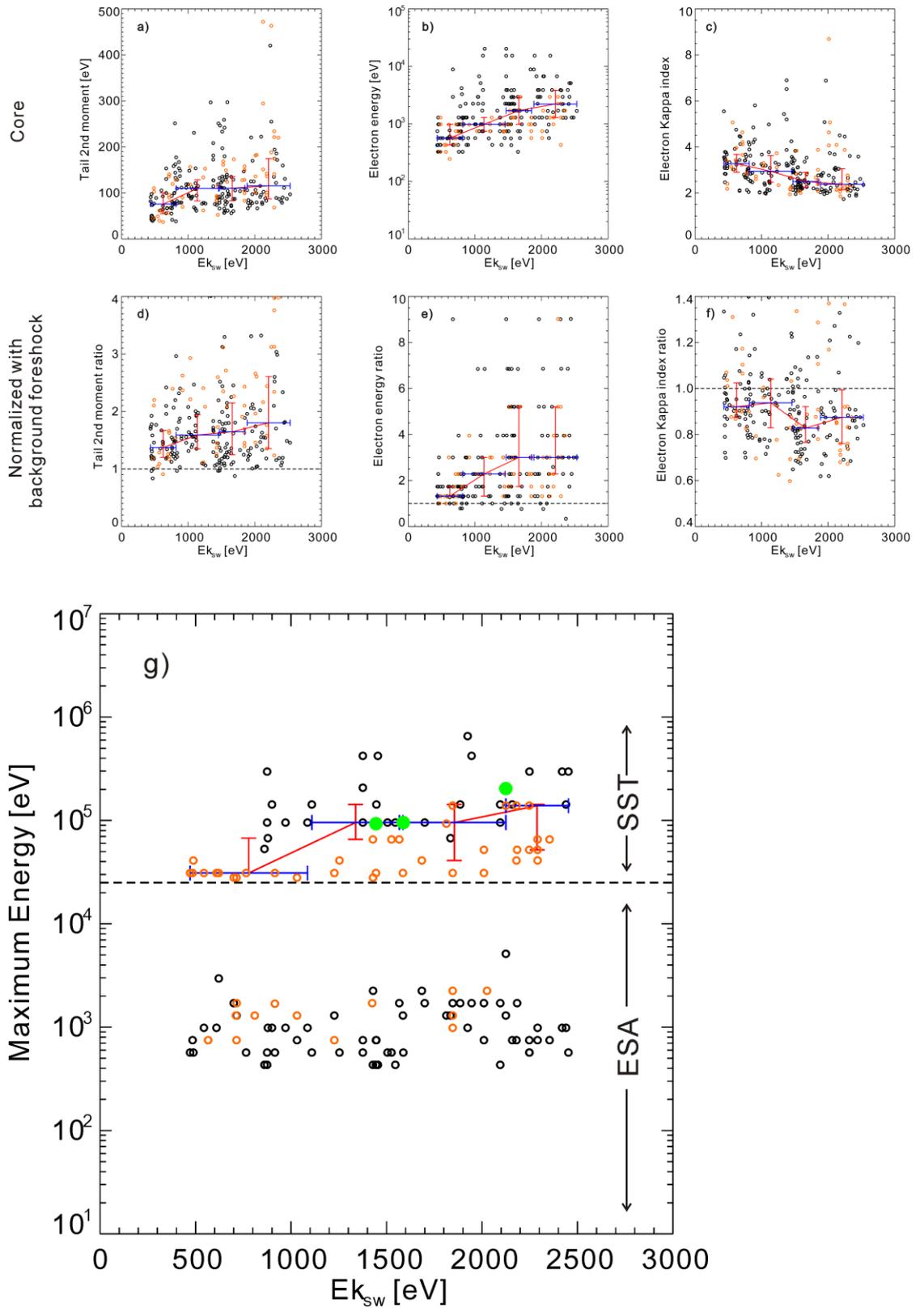

Figure 4. Electron energization estimates plotted against solar wind kinetic energy: average temperature of suprathermal tail (a); $E_{5\%}$, the energy above which the bins contribute 5% to the partial pressure (b); Kappa index of the phase space density distribution (c); and normalized estimates for these quantities (d, e, f, respectively) are their ratios to the background. Dashed horizontal lines in (d, e, f) are at ratio = 1, which indicates that most events have electrons more energized than those in the background foreshock. Electron energy energization estimates with and without normalization to the background are all positively correlated with the solar wind speed. The maximum energy that can be detected by particle instruments, another measure of energization, is shown in (g). Higher maximum electron energy is more likely seen at higher solar wind speed. Green dots in (g) indicate the events reported in Wilson et al. [2016]. Red lines are medians and error bars signify quartiles (25% and 75%). Dark blue horizontal bars indicate the data bins in which median and error bars are calculated (each data bin contains equal data number). Orange indicates events in Particle Burst mode. Events in Particle Burst mode show no statistical differences from events in Fast Survey mode (black) in (a)-(f). In (g), only orange circles can be seen around 25 keV, because we require that events in Fast Survey mode be detected by at least the three lowest energy channels of SST.

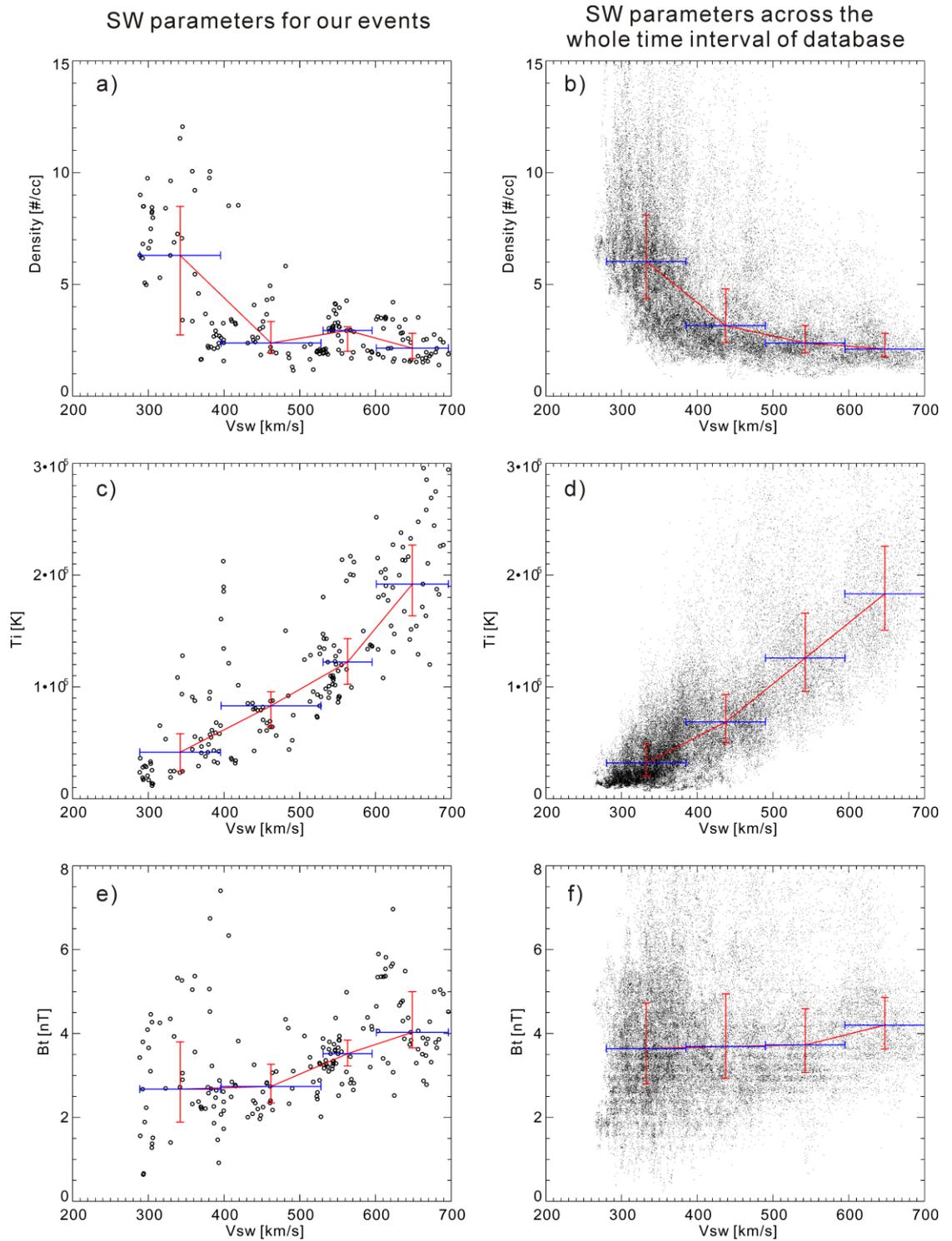

Figure 5. Solar wind density, temperature, and field strength versus the solar wind speed for our foreshock transients (left column) and in the ambient solar wind in the time period encompassing our database. Red lines are medians and error bars correspond to quartiles (25% and 75%). Dark blue horizontal bars indicate the data

bins in which median and error bars are calculated (for events, each data bin contains equal data number; for ambient solar wind, each data bin contains equal solar wind speed interval). The relationship between solar wind density and solar wind speed, and between solar wind temperature and solar wind speed of our events are similar to those from the entire time period of the database. But the solar wind magnetic field strength in our events (e) is almost along the lower bound of the solar wind magnetic field from the entire time period of the database (f).

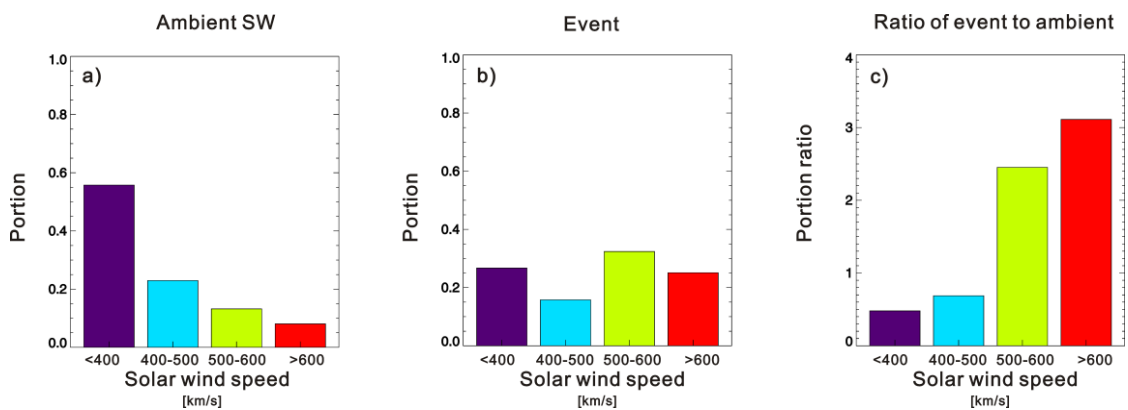

Figure 6. Solar wind speed probability distribution for the ambient solar wind encompassing our database (a) and for our events (b). The ratio of distribution for our events to that for the entire database is also shown to facilitate direct comparison (c). The probability of foreshock transient formation increases with the solar wind speed.

**Acknowledgment**

We thank the THEMIS software team and NASA's Coordinated Data Analysis Web (CDAWeb, http://cdaweb.gsfc.nasa.gov/) for their analysis tools and data access. Work supported by NASA contract NAS5-02099. The THEMIS data and THEMIS software (TDAS, a SPEDAS.org plugin) are available at http://themis.ssl.berkeley.edu. OMNI data used in this paper for the solar wind bulk velocity, density, temperature, and magnetic field can be found in the CDAWeb (http://cdaweb.gsfc.nasa.gov/).

Frame transform software are available at https://github.com/lynnbwilsoniii/wind_3dp_pros/wiki.